\documentclass{PoS}

\usepackage{amsmath,amssymb}
\usepackage{graphicx}

\title{Hybrid Hadronization}

\ShortTitle{Hybrid Hadronization}

\author{\speaker{Rainer J.\ Fries}
          \\
        Cyclotron Institute and Department of Physics and Astronomy, Texas A\&M University, College Station TX 77843, USA\\
        E-mail: \email{rjfries@comp.tamu.edu}}

\author{Michael Kordell\\
       Cyclotron Institute, Texas A\&M University, College Station TX 77843, USA\\
        E-mail: \email{mkordell@tamu.edu}}

\abstract{We discuss Hybrid Hadronization, a hadronization model which interpolates between string fragmentation in dilute parton systems and quark recombination in dense parton systems. We lay out the basic principles, discuss some details 
of the implementation, and show some prelimiary results. Hybrid Hadronization is realized as a software package which 
works with PYTHIA 8 and will be released publicly in the near future.}

\FullConference{International Conference on Hard and Electromagnetic Probes of High-Energy Nuclear Collisions\\
		30 September - 5 October 2018\\
		Aix-Les-Bains, Savoie, France}

\begin{document}


Hadronization of partons into hadrons is poorly understood from first principles. Yet most observables in high energy
collisions with hadrons in the final state are to some degree sensitive to hadronization 
dynamics. Over the past decades phenomenological models of hadronization have been developed. They can be tuned to describe a large amount of data consistently. Lund string fragmentation \cite{Andersson:1983ia} and quark recombination \cite{Fries:2008hs} are among the most successful hadronization models available.
Hybrid hadronization is proposed as a new hadronization model which incorporates both Lund string
fragmentation and quark recombination. It interpolates smoothly between them based on physics principles discussed
below. It recovers string fragmentation and pure quark recombination in limits of very dilute and very dense parton
systems, respectively.

Lund string fragmentation has been developed to resemble properties of the QCD vacuum in which gluon field lines tend to 
create flux tubes over long distances. In Monte Carlo generators these QCD strings can be used to connect color charges 
over large distances to make color singlet structures which then break up into hadrons 
String formation and fragmentation have been incorporated into the PYTHIA 8 event generator \cite{Sjostrand:2014zea} . 
This model has been extremely successful in describing hadronization in dilute sytems like $e^++e^- \to \text{hadrons}$ and 
$p+p$.
Quark recombination had first been considered in the 1970s and has seen a revival with the advent of data from high 
energy heavy ion colliders RHIC and LHC.  It was proposed that in a phase space filled with partons, the valence quarks 
of hadrons do not have to be created by string fragmentation, but are simply taken from the available quarks and 
antiquarks. Several key observables, like enhanced baryon production in A+A and the constituent quark scaling 
of elliptic flow $v_2$ can be naturally explained through quark recombination \cite{Fries:2008hs}.
The Hybrid Hadronization code has been created as part of the JETSCAPE effort \cite{Kauder:2018cdt}.
Its goal is to describe both vacuum and in-medium hadronization consistently. Hybrid Hadronization should eventually be
applicable to $p+p$, $p+$A and A$+$A systems alike.

We propose the following picture. The phenomenological $q\bar q$ potential in QCD
is characterized by a Coulomb-like part at small distances and a linear part at large distances.
The latter gives rise to QCD strings while 
at small distances the potential allows the direct formation of 
bound states.
Thus in a given system of quarks one should loop over available quark-antiquark pairs and triplets of quarks and antiquarks, 
and calculate their probability to directly form a bound state or resonance. If recombination fails for a quark it has to rather 
form a string with a partner. 
This procedure will turn a system of quarks into a system of 
recombined hadrons and resonances plus QCD strings. The strings capture all partons with large separation from other colored 
objects. The length scale is set by the typical size of hadron wave functions. Strings can then be handed over to PYTHIA for fragmentation.
The quark recombination step is suitable to describe medium effects for perturbative parton showers. Sampled thermal partons can simply be added to partons from a shower Monte Carlo. All partons are subsequently processed equally, allowing for recombination of mesons and baryons that contain both thermal and shower partons.

The quark recombination algorithm used here was first proposed in \cite{Han:2016uhh} based on the instantaneous recombination model
\cite{Fries:2008hs}. It uses the quark model and hadron wave functions based on harmonic oscillator potentials for simplicity.
The widths of the wave functions for ground states have been fitted to measured charge radii where available. 
The recombination probability for two wave packets, representing quark and antiquark, and assumed to be Gaussian
for simplicity, can then be calculated in the Wigner formalism. The Wigner formalism is convenient to use information
available in both space and momentum space. For example, for light quarks and antiquarks with wave packets
centered around phase space coordinates $(\mathbf{x}_1,\mathbf{p}_1)$ and $(\mathbf{x}_2,\mathbf{p}_2)$,
respectively, the recombination probability into a ground state meson with wave function
width $\sigma$, at a particular time $t$, is
\begin{equation}
  P(u) = e^{-u}  \quad \text{where}\quad u = \frac{1}{2 \sigma^2} \left( \mathbf{x}_2-\mathbf{x}_1\right)^2
  + \frac{\sigma^2}{8} \left( \mathbf{p}_2-\mathbf{p}_1\right)^2
\end{equation}
if the quark wave packets are assumed to have width $\sigma/\sqrt{2}$. The latter is chosen for convenience.\footnote{Monte Carlo generators usually do not provide realistic estimates for widths of wave packets.}
All probabilities have to be evaluated in the common rest frame of the quark-antiquark pair \cite{Han:2016uhh}.
Spins of partons are treated purely statistical. 
Color can be treated in two different ways. Either purely statistical, which is a good approximation in a sample of
mostly thermal partons. Or color can be traced through color tags, as used in PYTHIA 8, which is desirable
in jet showers in $p+p$ collisions and other dilute systems \cite{kordellfries}.

The Hybrid Hadronization code is written in C++ and needs PYTHIA 8 to run. Internally it consists of four different stages. 
\begin{itemize}
\item[(I)] Read in information about partons to be hadronized. The code accepts individual parton jet showers from
  Monte Carlo codes like PYTHIA or MATTER \cite{Majumder:2013re}. It also acceptes full events, like PYTHIA $p+p$ events.
  String structures and color tags, if available, are preserved. Thermal partons can be added at this stage, 
  e.g.\ by sampling the $T=T_c$ hypersurface
  from a fluid dynamic simulation of A+A collisions. Gluons are always split into quark-antiquark 
  pairs.
\item[(II)] All combinations of parton pairs and triplets are tested for recombination. For each the probability
  from color$\times$spin$\times$wave function overlap as described above is evaluated. A die role determines if a 
  recombination under consideration actually happens, and into which of the available hadron or hadron resonance channels it
  proceeds. The formation of hadrons from thermal partons only is currently disabled. All other hadrons formed during 
  this stage are written to the Hybrid Hadronization output, including resonances. 
\item[(III)] As partons from strings recombine, holes will appear in the original QCD strings. They are repaired according 
  to an algorithm which will be described in detail elsewhere \cite{kordellfries}. As an example, the simplest case of string repair happens when a quark or antiquark terminating a string has recombined with a thermal parton, in which case the missing string parton is replaced 
  by a suitable thermal parton.
\item[(IV)] In the last step all surviving string structures are surveyed if they can be handled by PYTHIA 8, and they are 
  processed further if needed. For example, the previous string repair step might have created systems with multiple 
  junctions. The output after step (IV) consists of a list of recombined hadrons and resonances, and the remnant string system. 
  It can now be handed over to PYTHIA 8 which will fragment the strings and decay all resonances according to user settings.
\end{itemize}

\begin{figure}[tb]
\begin{center}
	\includegraphics[width=0.7\columnwidth]{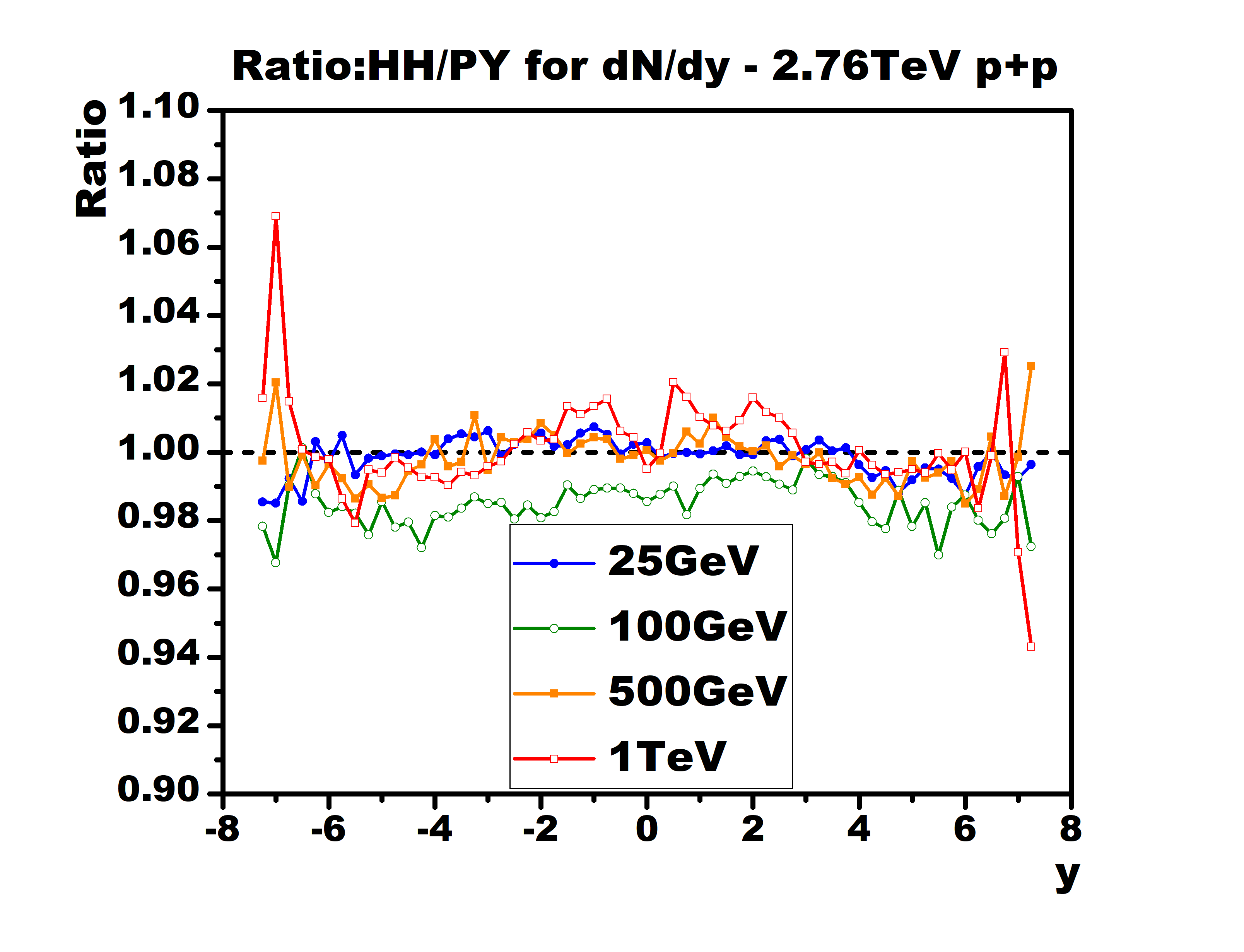}
\end{center}
  \caption{A preliminary result obtained from Hybrid Hadronization: The ratio of charged hadron yields in $p+p$ 
  collisions at $\sqrt{s}=2.76$ TeV hadronized with Hybrid Hadronization and PYTHIA 8, as a function of hadron rapidity $y$
  and for different $\hat p_T$ (25 GeV, 100 GeV, 500 GeV, 1 TeV) in PYTHIA 8. Deviations between the two calculations
  are generally less than 2 \%. }
  \label{fig:1}
\end{figure}

\begin{figure}[tb]
\begin{center}
   \includegraphics[width=0.7\columnwidth]{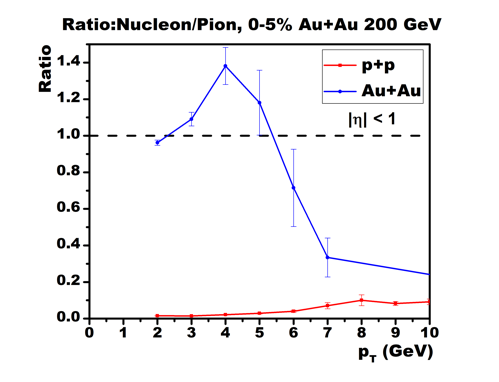}
\end{center}
  \caption{A preliminary result obtained from Hybrid Hadronization: The proton/pion ratio in Au+Au collisions at $\sqrt{s_\text{NN}}=200$ GeV and $p+p$ collisions
  at the same energy. Thermal-shower recombination is allowed in the Au+Au case but purely thermal bulk hadrons are not
  included. Thermal-shower recombination leads to the signature enhancement of baryons in the Au+Au case.}
  \label{fig:2}
\end{figure}

Testing of the Hybrid Hadronization code is ongoing but we can report preliminary results. Note that not tuning to data has been 
carried out.
To restate the goals,
we would like to interfere minimally with the success story of Lund strings in dilute systems. At the same time, we would like
to reproduce signature effects of recombination in A+A collisions. The likelihood of quark recombination in jet showers
has already been studied to some extent in \cite{Han:2016uhh} and
we refer the reader to that reference for more details. The conclusion is that jets seem to naturally feature a bulk region
in which several partons can be quite close together in phase space, in addition to a very dilute tail. 
The bulk part is associated with the low-$z$ part of the shower, where
$z$ as usual is the momentum fraction in a jet, while the large-$z$ part is typically dilute. Note that this is a statement
averaged over many showers. Fluctuations between showers are very large due to the relatively small number of
partons.

Fig.\ \ref{fig:1} shows the ratio of rapitidy spectra in minimum bias $p+p$ collisions at 2.76 TeV obtained from Hybrid Hadronization and pure PYTHIA 8 string fragmentation. Full PYTHIA 8 parton events have 
been used in both cases as input to hadronization. The four different lines correspond to events reported by PYTHIA with an 
initial momentum transfer $\hat p_T$ in a bin around the value shown. Generally we find rather small deviations between 
Hybrid Hadronization and pure string fragmentation in minimum bias $p+p$.
Fig.\ \ref{fig:2} shows the ratio of protons vs pions as a function of transverse momentum $P_T$ around midrapidity
for Au+Au collisions at 200 GeV. Enhanced baryon production has been a hallmark of quark recombination models, and
we see this behavior reproduced by Hybrid Hadronization. Here individual jet showers created by MATTER \cite{Majumder:2013re}
were embedded into fluid dynamic events. The resulting enhancement can be traced back to shower-thermal recombination
into baryons. They are compared to MATTER showers in $p+p$. Note that no underlying event in $p+p$ and no thermal
hadrons in A$+$A have been added, thus results below $P_T=2$ GeV/$c$ are not shown.

To summarize, Hybrid Hadronization is a new hadronization model which combines features of 
string fragmentation and quark recombination. The goal is a
hadronization model that can describe the transition between very dilute systems like $p+p$ and even $e^++e^-$,
and heavy ion collisions. The code is currently undergoing testing and will be publicly released
in a future JETSCAPE version. 

This work was supported by the US National Science Foundation under award nos.\ 1516590, 1550221 and 1812431.


\begin{thebibliography}{99}

\bibitem{Andersson:1983ia} 
  B.~Andersson {\it et al.},
  Phys.\ Rept.\  {\bf 97}, 31 (1983).

\bibitem{Fries:2008hs} 
  R.~J.~Fries, V.~Greco and P.~Sorensen,
  Ann.\ Rev.\ Nucl.\ Part.\ Sci.\  {\bf 58}, 177 (2008).

\bibitem{Sjostrand:2014zea} 
  T.~Sj\"ostrand {\it et al.},
  Comput.\ Phys.\ Commun.\  {\bf 191}, 159 (2015).

\bibitem{Kauder:2018cdt} 
  K.~Kauder [JETSCAPE Collaboration],
  arXiv:1807.09615 [hep-ph].

\bibitem{Han:2016uhh} 
  K.~C.~Han, R.~J.~Fries and C.~M.~Ko,
  Phys.\ Rev.\ C {\bf 93}, 045207 (2016).

\bibitem{kordellfries} M.\ Kordell and R.\ J.\ Fries, {\it to be published}.

\bibitem{Majumder:2013re} 
  A.~Majumder,
  Phys.\ Rev.\ C {\bf 88}, 014909 (2013).



\end{thebibliography}
\end{document}